\documentclass[iicol,pdflatex,sn-mathphys-num]{sn-jnl}

\usepackage{graphicx}%
\usepackage{multirow}%
\usepackage{amsmath,amssymb,amsfonts}%
\usepackage{amsthm}%
\usepackage{mathrsfs}%
\usepackage[title]{appendix}%
\usepackage{xcolor}%
\usepackage{textcomp}%
\usepackage{manyfoot}%
\usepackage{booktabs}%
\usepackage{algorithm}%
\usepackage{algorithmicx}%
\usepackage{algpseudocode}%
\usepackage{listings}%
\usepackage[acronym]{glossaries}%
\usepackage{subcaption}
\usepackage{multirow}
\usepackage{placeins}
\usepackage{xurl}
\let\orcidlogo\relax
\usepackage{orcidlink}

\theoremstyle{thmstyleone}%
\theoremstyle{thmstyletwo}%

\theoremstyle{thmstylethree}%
\makeglossaries

\newacronym{uboone}{MicroBooNE}{Micro Booster Neutrino Experiment}
\newacronym{mboone}{MiniBooNE}{Mini Booster Neutrino Experiment}
\newacronym{sbn}{SBN}{Short Baseline Neutrino}
\newacronym{sbnd}{SBND}{Short Baseline Near Detector}
\newacronym{dune}{DUNE}{Deep Underground Neutrino Experiment}
\newacronym{fnal}{FNAL}{Fermi National Accelerator Laboratory}
\newacronym{lbnf}{LBNF}{Long-Baseline Neutrino Facility}
\newacronym{nd}{ND}{Near Detector}
\newacronym{fd}{FD}{Far Detector}
\newacronym{surf}{SURF}{Sanford Underground Research Facility}
\newacronym{cern}{CERN}{Conseil Européen pour la Recherche Nucléaire}
\newacronym{lsnd}{LSND}{Liquid Scintillator Neutrino Detector}
\newacronym{sk}{Super-K}{Super-Kamiokande}
\newacronym{sno}{SNO}{Sudbury Neutrino Observatory}
\newacronym{mi}{MI}{Main Injector}
\newacronym{hk}{Hyper-K}{Hyper-Kamiokande}
\newacronym{gallex}{GALLEX}{Gallium Experiment}
\newacronym{sage}{SAGE}{Soviet-American Gallium Experiment}
\newacronym{sbl}{SBL}{Short-Baseline}
\newacronym{jparc}{JPARC}{Japan Proton Accelerator Research Complex}
\newacronym{iwcd}{IWCD}{Intermediate Water Cherenkov Detector}
\newacronym{reno}{RENO}{Reactor Experiment for Neutrino Oscillation}
\newacronym{neos}{NEOS}{Neutrino Experiment for Oscillation at Short Baseline}
\newacronym{lariat}{LArIAT}{Liquid Argon In A Testbeam}
\newacronym{argoneut}{ArgoNeuT}{Argon Neutrino Teststand}
\newacronym{icarus}{ICARUS}{Imaging Cosmic And Rare Underground Signals}

\newacronym{lar}{LAr}{liquid argon}
\newacronym{tpc}{TPC}{Time Projection Chamber}
\newacronym{lartpc}{LArTPC}{Liquid Argon Time Projection Chamber}
\newacronym{lcs}{LCS}{Light Collection System}
\newacronym{crt}{CRT}{Cosmic-Ray Tagger}
\newacronym{apa}{APA}{Anode Plane Assembly}
\newacronym{cpa}{CPA}{Cathode Plane Assembly}
\newacronym{pmt}{PMT}{Photomultiplier Tube}
\newacronym{tpb}{TPB}{Tetraphenyl butadiene}
\newacronym{asic}{ASIC}{application-specific integrated circuit}
\newacronym{daq}{DAQ}{Data Acquisition}
\newacronym{adc}{ADC}{Analogue-to-Digital Converter}
\newacronym{fem}{FEM}{Front End Module}
\newacronym{fpga}{FPGA}{Field-Programmable Gate Array}
\newacronym{tb}{TB}{Trigger Board}
\newacronym{sp}{SP}{Single Phase}
\newacronym{dp}{DP}{Dual Phase}
\newacronym{hd}{HD}{Horizontal Drift}
\newacronym{vd}{VD}{Vertical Drift}
\newacronym{ce}{CE}{Cold Electronics}
\newacronym{pds}{PDS}{Photon Detection System}
\newacronym{fc}{FC}{Field Cage}
\newacronym{arapuca}{ARAPUCA}{Argon R\&D Advanced Program at UniCAmp}
\newacronym{sipm}{SiPM}{silicon photomultiplier}
\newacronym{lhcb}{LHCb}{Large Hadron Collider beauty}

\newacronym{bnb}{BNB}{Booster Neutrino Beam}
\newacronym{numi}{NuMI}{Neutrinos at the Main Injector}

\newacronym{pot}{POT}{Protons-On-Target}
\newacronym{vuv}{VUV}{Vacuum Ultraviolet}
\newacronym{tbp}{TB}{Test Beam}
\newacronym{cla}{CL}{Confidence Level}
\newacronym{uav}{UAV}{unmanned aerial vehicle}

\newacronym{sce}{SCE}{Space-Charge Effect}
\newacronym{edm}{EDM}{Event Data Model}
\newacronym{pfo}{PFO}{Particle-Flow Object}

\newacronym{sm}{SM}{Standard Model}
\newacronym{bsm}{BSM}{Beyond the Standard Model}
\newacronym{rh}{RH}{right-handed}
\newacronym{lh}{LH}{left-handed}
\newacronym{cp}{CP}{Charge-Parity}
\newacronym{ssm}{SSM}{Standard Solar Model}
\newacronym{pmns}{PMNS}{Pontecorvo–Maki–Nakagawa–Sakata}
\newacronym{msw}{MSW}{Mikheyev-Smirnov-Wolfenstein}
\newacronym{no}{NO}{Normal Ordering}
\newacronym{io}{IO}{Inverted Ordering}
\newacronym{nh}{NH}{Normal Hierarchy}
\newacronym{ih}{IH}{Inverted Hierarchy}
\newacronym{cc}{CC}{charged-current}
\newacronym{nc}{NC}{neutral current}
\newacronym{es}{ES}{Elastic Scattering}
\newacronym{qe}{QE}{Quasi-Elastic}
\newacronym{el}{El}{Elastic}
\newacronym[description={One-particle one-hole}]{1p1h}{1p-1h}{one-particle one-hole}
\newacronym[description={Two-particle two-hole}]{2p2h}{2p-2h}{two-particle two-hole}
\newacronym{mec}{MEC}{Meson Exchange Currents}
\newacronym{coh}{Coh}{Coherent}
\newacronym{res}{Res}{Resonant}
\newacronym{dis}{DIS}{Deep Inelastic Scattering}
\newacronym{ia}{IA}{Impulse Approximation}
\newacronym{rfg}{RFG}{Relativistic Fermi-Gas}
\newacronym{lfg}{LFG}{Local Fermi-Gas}
\newacronym{fsi}{FSI}{Final State Interactions}
\newacronym{sint}{SI}{Secondary Interactions}
\newacronym{ckm}{CKM}{Cabibbo-Kobayashi-Maskawa}
\newacronym{gut}{GUTs}{Grand Unified Theories}

\newacronym{mip}{MIP}{Minimum Ionising Particle}
\newacronym{mcs}{MCS}{Multiple Coulomb Scattering}
\newacronym{ly}{LY}{Light-Yield}
\newacronym{pe}{PE}{Photoelectron}

\newacronym{qed}{QED}{Quantum Electrodynamics}
\newacronym{qcd}{QCD}{Quantum Chromodynamics}
\newacronym{hnl}{HNL}{Heavy Neutral Lepton}
\newacronym{qm}{QM}{Quantum Mechanics}
\newacronym{snu}{SNU}{Solar Neutrino Unit}
\newacronym{si}{SI}{Système International}

\newacronym[description={Muon-neutrino charged-current single charged pion}]{cc1pi}{CC1$\pi^{\pm}$}{muon-neutrino charged-current single charged pion}
\newacronym{fv}{FV}{Fiducial Volume}
\newacronym{pdg}{PDG}{Particle Data Group}

\newacronym{fkr}{FKR}{Feynman-Kislinger-Ravndal}
\newacronym[description={Cosmic Ray}]{cr}{CR}{cosmic ray}
\newacronym{fom}{FoM}{Figure of Merit}
\newacronym{bdt}{BDT}{Boosted Decision Tree}
\newacronym{mc}{MC}{Monte-Carlo}
\newacronym{sw}{SW}{Sanford-Wang}
\newacronym{cv}{CV}{Central-Value}
\newacronym{pg}{PG}{Probability-of-Goodness}
\newacronym{red}{R$\&$D}{Research and Development}
\newacronym{ad}{AD}{Antineutrino Detector}
\newacronym{iw}{IW}{Inner Water}
\newacronym{ow}{OW}{Outer Water}
\newacronym{rpc}{RPC}{Resistive Plate Chamber}

\newacronym{ai}{AI}{artificial intelligence}
\newacronym{ml}{ML}{machine learning}
\newacronym{dl}{DL}{deep learning}
\newacronym{ann}{ANN}{artificial neural network}
\newacronym{cnn}{CNN}{convolutional neural network}
\newacronym{cpu}{CPU}{central processing unit}
\newacronym{gpu}{GPU}{graphics processing unit}
\newacronym{tpu}{TPU}{tensor processing unit}
\newacronym{tf}{Tf}{TensorFlow}
\newacronym{iot}{IoT}{Internet-of-Things}
\newacronym{ram}{RAM}{Random Access Memory}
\newacronym{api}{API}{Application Programming Interface}
\newacronym{qat}{QAT}{quantisation aware training}
\newacronym{tdp}{TDP}{thermal design power}
\newacronym{vram}{VRAM}{Video Random Access Memory}
\newacronym{hz}{Hz}{hertz}
\newacronym{bios}{BIOS}{Basic Input/Output System}
\newacronym{rom}{ROM}{Read-Only Memory}
\newacronym{alu}{ALU}{arithmetic logic unit}
\newacronym{fpu}{FPU}{Floating Point Unit}
\newacronym{mac}{MAC}{Multiply-Accumulate}
\newacronym{mxu}{MXU}{Matrix Multiplication Unit}
\newacronym{hbm}{HBM}{High Bandwidth Memory}
\newacronym{npu}{NPU}{Neural Processing Unit}
\newacronym{mosfet}{MOSFET}{metal–oxide–semiconductor field-effect 
transistors}
\newacronym{tmva}{TMVA}{Toolkit for Multivariate Data Analysis}
\newacronym{mva}{MVA}{Multivariate Data Analysis}
\newacronym{resnet}{ResNet}{Residual Neural Network}
\newacronym{dense}{DenseNet}{Densely Connected Convolutional Network}
\newacronym{ptiq}{PTIQ}{post-training integer quantisation}
\newacronym{ptq}{PTQ}{post-training quantisation}
\newacronym{roc}{ROC}{Receiver Operating Characteristic}
\newacronym{tops}{trillion operations per second}{TOPS}

\newacronym{pmf}{pmf}{probability mass function}
\newacronym{cdf}{cdf}{cumulative distribution function}
\newacronym{pdf}{pdf}{probability density function}
\newacronym{clt}{CLT}{central limit theorem}
\newacronym{pca}{PCA}{Principal Component Analysis}
\newacronym{svm}{SVM}{Support Vector Machine}
\newacronym{lda}{LDA}{Linear Discriminant Analysis}
\newacronym{mle}{MLE}{Maximum Likelihood Estimation}
\newacronym{sgd}{SGD}{Stochastic Gradient Descent}
\newacronym{dof}{dof}{degrees of freedom}

\newacronym{tki}{TKI}{Transverse Kinematic Imbalance}
\newacronym{tmi}{TMI}{Transverse Momentum Imbalance}
\newacronym{tba}{TBA}{Transverse Boosting Angle}
\newacronym{aba}{ABA}{Azimuthal Boosting Angle}
\newacronym{cl}{CL}{Collaborative Learning}
\newacronym{hep}{HEP}{High Energy Physics}

\newacronym{mit}{MIT}{Massachusetts Institute of Technology}

\newacronym{su}{SU}{Special Unitary}
\newacronym{ssb}{SSB}{Spontaneous symmetry breaking}

\begin{document}

\hypersetup{
    colorlinks=true,
    linkcolor=black,
    citecolor=black,
    urlcolor=black,
    filecolor=black
}

\title[Article Title]{Physics at the Edge: Benchmarking Quantisation Techniques and the Edge TPU for Neutrino Interaction Recognition}

\author*[1,2]{\fnm{Stefano} \sur{Vergani}\,\orcidlink{0000-0001-8653-3841}}\email{stefano.vergani@kcl.ac.uk}

\author[3]{\fnm{Hilary} \sur{Utaegbulam}\,\orcidlink{0000-0003-4583-4546}}

\author[4]{\fnm{Michael} \sur{Wang}\,\orcidlink{0000-0002-4713-9646}}

\author[2]{\fnm{Leigh H.} \sur{Whitehead}\,\orcidlink{0000-0002-3327-2534}}

\author[1]{\fnm{Arden S. H.} \sur{Tsang}\,\orcidlink{0009-0008-0782-5394}}

\author[5]{\fnm{Lorenzo} \sur{Uboldi}\,\orcidlink{0000-0002-6969-4894}}

\affil[1]{\orgdiv{Department of Physics}, \orgname{King's College London}, \orgaddress{\city{London}, \country{UK}}}

\affil[2]{\orgdiv{Cavendish Laboratory}, \orgname{University of Cambridge}, \orgaddress{\city{Cambridge}, \country{UK}}}

\affil[3]{\orgdiv{Department of Physics and Astronomy}, \orgname{University of Rochester}, \orgaddress{\city{Rochester}, \state{NY}, \country{USA}}}

\affil[4]{\orgname{Fermi National Accelerator Laboratory}, \orgaddress{\city{Batavia}, \state{IL}, \country{USA}}}

\affil[5]{\orgdiv{Department of Physics}, \orgname{Politecnico di Milano}, \orgaddress{\city{Milan}, \country{Italy}}}

\abstract{This work presents a benchmark of quantisation techniques for convolutional neural networks applied to neutrino interaction recognition, motivated by the need for low-power inference close to the detector in future liquid argon time-projection chamber (LArTPC) experiments. Using simulation of a generic LArTPC, four Keras models are trained to classify neutral-current, muon-neutrino charged-current and electron-neutrino charged-current interactions, quantised using post-training integer quantisation and quantisation-aware training, and deployed on the Google Coral Edge TPU. The accuracy degradation introduced by quantisation and deployment is strongly model-dependent: InceptionV3 shows no significant loss along either pipeline, whereas the other models are substantially degraded. Latency and energy per event on the Edge TPU are then compared with those of an Intel Core i7-13650HX CPU, a consumer NVIDIA RTX 4060 GPU and a data-centre NVIDIA A100 GPU, using measured energy at different batch sizes and thread counts; the energy of the Edge TPU is estimated from its nominal power. For InceptionV3 at batch size 1, the Edge TPU is 1.3 times faster than the CPU and 4.7--4.9 times slower than the GPUs, while requiring 6--34 times less energy per event than the other devices (5--70 times across all four models); with batching, the GPUs reduce this gap to a factor of 2.7--2.8. The Edge TPU is also approximately four times cheaper than the consumer GPU and more than two orders of magnitude cheaper than the A100. Its low power per device eases deployment where power and cooling are constrained, with an additional benefit in terms of environmental impact. This paper explores possible future integrations of edge AI technologies with neutrino physics.}

\keywords{Edge TPU, Deep Learning, Neutrinos, Quantisation, AI, Edge AI, Google Coral}



\maketitle

\section{Introduction}\label{sec:intro}

Over the last decade, \gls{ai} has played a revolutionary role in science, including particle physics \cite{Schwartz2021Modern, BhattacherjeeMukherjee2024}. In particular, advanced \glspl{cnn} have reshaped the way pattern recognition is performed in experiments, allowing for high accuracy \cite{liu2020deeplearningbasedkinematicreconstructiondune} and low latency\footnote{In this work, the term latency refers to end-to-end inference time per sample, not communication delay. Latency and speed are used interchangeably throughout this paper.}\cite{Abratenko2026MicroBooNE}. This includes \gls{ai} for fast triggering, where rare signatures need to be detected in very short timescales \cite{UBOLDI2022166371, Albrecht2025TriggerLHC}. Traditionally, advanced \gls{ai} algorithms have been trained and deployed on \glspl{gpu}. They provide great performance in terms of speed \cite{wang2019benchmarkingtpugpucpu}, but \glspl{gpu} are expensive and have a high power consumption \cite{Godoy_2025, falk2025flopsfootprintsresourcecost}. This translates into greater electrical power required by the device and the cooling system, and unsustainable environmental heating. The impact of \gls{ai} on the environment is considered one of the biggest issues of this decade \cite{ wegmeth2025greenrecommendersystemsunderstanding, Pathania_2025, ALNAFRAH2025126813, Wang2024AIenvironment, elsworth2025measuringenvironmentalimpactdelivering, falk2025carboncradletograveenvironmentalimpacts, Nik2025DecodingEnergy}, and solving this remains on the priority list of many strategic white papers \cite{Elvira2024SnowmassComputing, Caron_2026}. Another issue is that \gls{gpu}s are rarely deployed close to the detector or source of data, but are often stored in datacentres away from it; this is the case for experiments at the \gls{fnal} \cite{savard2023optimizinghighthroughputinference} and the \gls{cern} \cite{Suarez_2025}. In collider experiments, the radiation and magnetic-field environment precludes commercial \gls{gpu}s on-detector, so GPU-based triggers run in surface farms, with near-detector uses limited to cases such as the NA62 \gls{gpu}-based L0 trigger \cite{ammendola2016gpubasedrealtimetriggeringna62}. A recent attempt to solve this problem has been to use radiation-tolerant \gls{fpga}s to perform \gls{ml} tasks on the \gls{lhcb} experiment \cite{Govorkova_2026}. In underground neutrino experiments, the radiation environment is benign, but the power, cooling and space available underground limit the use of \gls{gpu} racks near the detector. \gls{fpga}s are the established low-power alternative and offer deterministic, sub-microsecond latency, and tools such as \texttt{hls4ml}~\cite{Govorkova_2026} have simplified the deployment of neural networks on them. However, \glspl{lartpc} event recognition requires deep \gls{cnn} operating on high-resolution images, and \gls{fpga}s resource constraints have so far restricted implementations for \glspl{lartpc} data selection to very shallow networks \cite{10.3389/frai.2022.855184}.

Edge AI accelerators offer a complementary approach: they trade latency for the ability to run large, standard \gls{cnn} architectures at low power \cite{MUHOZA2023100930, tu2023deepen2023energydatasetsedge} with minimal hardware-specific development. Examples of their usage can be found in satellites \cite{Shi_2025}, \gls{uav} \cite{girgin2025edgeaidroneautonomousconstruction}, and even medicine \cite{Prabha2026EdgeAIHealthcare}. Physicists have also recently started to explore the potential of such devices for future detectors~\cite{gonski2026machinelearningheterogeneousedge}.
Edge \gls{ai}, broadly, denotes inference performed at or near the data source under tight power, latency and space constraints, and in particle physics it has so far been realised mainly on \glspl{fpga}~\cite{Duarte2018hls4ml, Govorkova_2026}. In this paper we use the term \emph{edge \gls{ai} device} in a narrower sense, to denote commercial off-the-shelf \glspl{asic} designed exclusively for low-precision neural-network inference at a power of a few watts, such as the Google Coral Edge \gls{tpu}. These differ from \glspl{fpga} in two respects: their silicon is fixed-function, so the user does not design the hardware, and they are programmed by compiling models from standard \gls{ml} frameworks rather than through hardware description languages or high-level synthesis. As a consequence, large, standard \gls{cnn} architectures can be deployed with quantisation as the only modification. \glspl{fpga} are not considered edge \gls{ai} devices in this sense.

This work presents the benchmark of various \glspl{cnn} running on the Edge \gls{tpu}, an edge \gls{ai} device produced by Google Coral, performing event recognition on a neutrino dataset simulating a \gls{lartpc}. The concept of the \gls{lartpc} was first presented in 1977 \cite{Rubbia:1977zz}, and since then it has been used in several successful neutrino experiments such as \gls{icarus} \cite{icaruscollaboration2023icarusfermilabshortbaselineneutrino}, \gls{argoneut} \cite{guenette2011argoneutexperiment}, \gls{lariat} \cite{Cavanna:2014iqa}, and \gls{uboone} \cite{MicroBooNE:2016pwy}. The setup of a \gls{lartpc} comprises a cryostat filled with \gls{lar} kept at 87\,K, at least one cathode and anode separated by an electric field, allowing ionisation electrons to drift towards the anode. The anode is typically instrumented with wire planes or pixel detectors. We have chosen to simulate a \gls{lartpc} detector because it represents the state-of-the-art technology and will continue to play a central role in next-generation neutrino detectors, such as \gls{dune} \cite{DUNE:2020jqi}. 

Future \gls{lartpc} experiments such as \gls{dune} will have to analyse an unprecedented amount of data \cite{10735136}. The AI deployed for live triggering and reconstruction tasks will need to excel on all three elements of the triad: accuracy, latency, and power consumption. Models need to have a high accuracy to recognise interactions and guarantee low systematic errors, key to delivering physics goals when dealing with rare interactions. Latency is crucial for triggering, but it is also a significant bottleneck for analysing data offline. Finally, power consumption is directly linked to the overall running costs of the experiment and the environmental pollution due to heat dissipation. GPUs historically perform well along the axes of accuracy and latency, but have significant issues with power consumption. This paper evaluates hardware designed to lower power consumption, and quantifies its impact on accuracy and latency.

\section{Edge TPU}\label{sec:edgetpu}

Edge computing has become increasingly useful for experiments where inference must occur close to the data source, and under strict power and latency budgets~\cite{Sun:2021}. The Google Coral Edge \gls{tpu}~\cite{seshadri2022evaluation} is an \gls{asic} designed specifically to accelerate machine learning inference; the ``Edge'' in its name refers to deployment at the edge, in the sense defined in Section~\ref{sec:intro}~\cite{Yazdanbakhsh:2021}.
The Edge \gls{tpu} is optimised for energy efficiency rather than peak throughput: it performs up to 4 \gls{tops} on 8-bit integer operands while consuming approximately 2\,W, i.e.\ 2\,\gls{tops}/W~\cite{Sun:2021,nutelescopetpu}. For comparison, Table~\ref{tab:devices} lists the characteristic performance and power figures of the \gls{cpu} and \gls{gpu} platforms considered in this work, whose power limits range from tens to hundreds of watts.

\begin{table*}[t]
\centering
\small
\setlength{\tabcolsep}{4pt}
\resizebox{\textwidth}{!}{%
\begin{tabular}{lccccc}
\toprule
\textbf{Device} & \textbf{Type} & \textbf{Processing units} & \textbf{Memory} & \textbf{Power [W]} & \begin{tabular}{@{}c@{}}\textbf{Approximate}\\\textbf{price [USD]}\end{tabular} \\
\midrule
Intel i7-13650HX       & CPU  & 14 cores, 20 threads   & Host DDR5          & 55--157 & 485 \\
NVIDIA RTX 4060 Laptop & GPU  & 3072 CUDA cores        & 8\,GB GDDR6        & 115     & 300 \\
NVIDIA A100 SXM4       & GPU  & 6912 CUDA + 432 tensor & 40\,GB HBM2        & 400     & 12{,}000--15{,}000 \\
Coral USB Accelerator  & ASIC & systolic array         & 8\,MB on-chip SRAM & $\sim$2 & 75 \\
\bottomrule
\end{tabular}%
}
\caption{Characteristic specifications, power figures and approximate prices of the devices considered~\cite{IntelI7, NVIDIARTX4060, NVIDIAA100, CoralDatasheet}.}
\label{tab:devices}
\end{table*}

The core architecture of the Edge \gls{tpu} relies on a systolic array design. Data flows through a grid of arithmetic logic units that are able to perform parallel matrix multiplications, the fundamental operation of \glspl{cnn}, without frequent access to memory~\cite{Yazdanbakhsh:2021}.

However, the Edge \gls{tpu} architecture imposes specific constraints on the neural networks it can support: it operates exclusively on 8-bit unsigned integers (\texttt{uint8}), whereas \gls{ml} models are typically trained in 32-bit floating point (\texttt{float32}). Models must therefore be quantised, using either \gls{ptq} or \gls{qat} (Section~\ref{sec:quantisation}), before deployment~\cite{Jacob:2017}.

\section{Neural Network Quantisation}\label{sec:quantisation}

Quantisation is the process of mapping a model’s floating-point weights and activations (e.g., \texttt{float32} or \texttt{float16}) to lower-precision numeric formats, e.g., \texttt{uint8}, to reduce memory footprint and computational cost. It is usually applied either via \gls{ptq} or \gls{qat}. Both techniques can be particularly valuable for deploying models on resource-constrained devices, such as edge devices, by enabling faster inference and lower power, typically at the cost of a small accuracy drop if properly configured \cite{zhang2023posttrainingquantizationneuralnetworks}.   

\subsection{Post-Training Quantisation}\label{sec:ptq}

\gls{ptq} converts a pre-trained floating-point model to a lower-precision representation without further training. The process involves a calibration step where a small representative dataset is passed through the model to characterise the distribution and range of the model's weights and activations. These statistics are used to determine the appropriate tensor and channel scaling factors and zero-point values for quantisation. The model's \texttt{float32} or \texttt{float16} weights are then converted to \texttt{uint8} or another low-precision format using these parameters.  

\subsection{Quantisation-Aware Training}\label{sec:qat}
\gls{qat} incorporates quantisation effects into training by inserting fake-quantisation operators into the network, which round weights and activations to the target low-precision grid in the forward pass while keeping them stored in \texttt{float32}. In the backward pass, gradients are computed in full precision, so that the model learns weights that minimise the loss under quantisation. \gls{qat} can be applied when training from scratch or, as in this work, as a fine-tuning step starting from a pre-trained floating-point model.

\section{Neutrino Dataset}\label{sec:datasets}
GENIE \texttt{v3\_00\_06}~\cite{Alam:2015nkk} was used to produce neutrino interactions with a uniform flux distribution in the range 1--4\,GeV, roughly approximating the main part of the DUNE neutrino flux distribution~\cite{DUNE:2020jqi}. Events were produced in three broad categories: \gls{cc} $\nu_\mu$ interactions, \gls{cc} $\nu_e$ interactions, and \gls{nc} interactions. The final-state particles produced in the neutrino interactions were passed through a simple LArTPC detector simulation~\cite{Chappell:2022yxd} based on Geant4 \texttt{v4\_10\_6}~\cite{Agostinelli:2002hh}. The simulated detector was a monolithic cuboid of liquid argon of dimensions $($x$,$y$,$z$) = (5\,\textrm{m}, 5\,\textrm{m}, 5\,\textrm{m})$, where $x$ was the electron drift direction, $y$ was the height, and $z$ was the neutrino beam direction. The 3D energy deposits produced by the ionising particles were projected onto three two-dimensional views ($u$, $v$, and $w$) in the $yz$-plane, similar to the readout planes planned for DUNE~\cite{DUNE:2020txw}. The planes were oriented at $\pm$35.9$^\circ$ and 0$^\circ$ to the vertical. Images of size 224$\times$224 pixels were produced for each of the readout views after cropping the images to cover an area of size 224\,cm$\times$224\,cm around the true interaction vertex, such that each pixel represented a 1\,cm$\times$1\,cm area of the readout plane. The sample of simulated neutrino interactions was initially used in Ref.~\cite{Vergani:2024syg}. A total of 22{,}338 neutrino interactions were used in this work, split into a training set of 17{,}338 events (CC $\nu_{\mu}$: 6{,}328, CC $\nu_{e}$: 4{,}742, NC: 6{,}268) and a test set of 5{,}000 events (CC $\nu_{\mu}$: 1{,}149, CC $\nu_{e}$: 2{,}696, NC: 1{,}155).

\begin{figure*}[htb]
    \centering
    \includegraphics[scale=0.26]{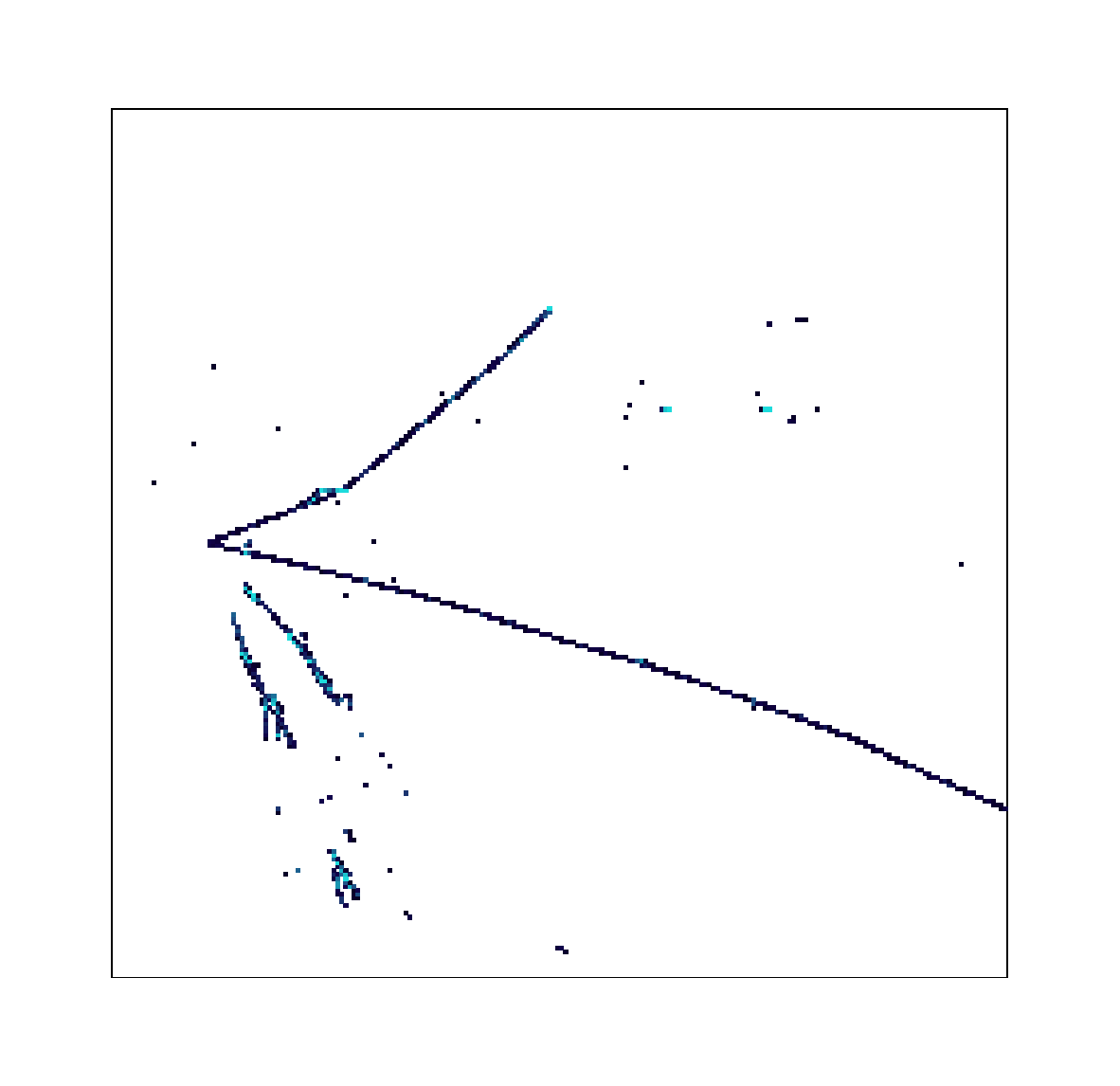}
    \includegraphics[scale=0.26]{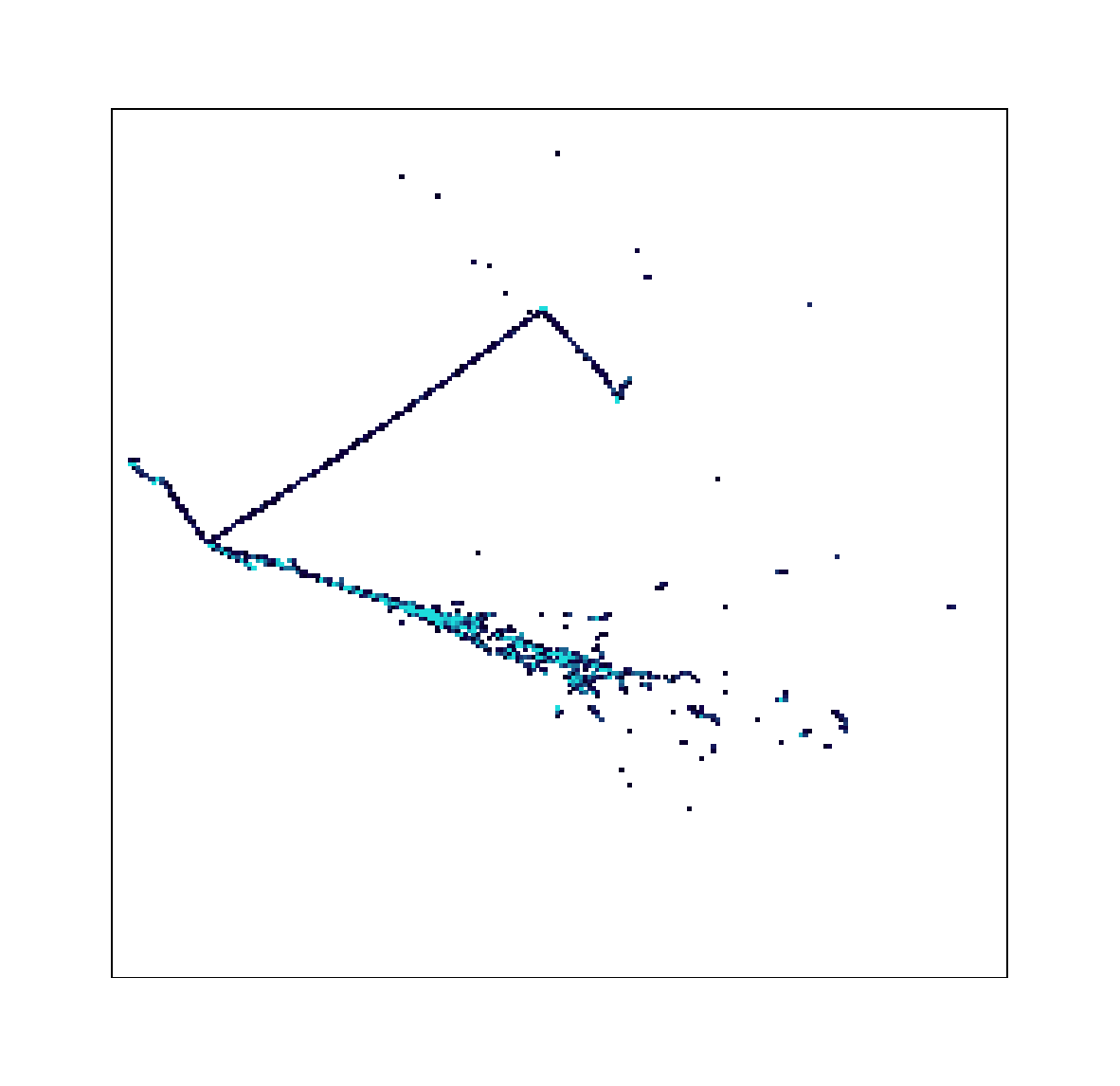}
    \includegraphics[scale=0.26]{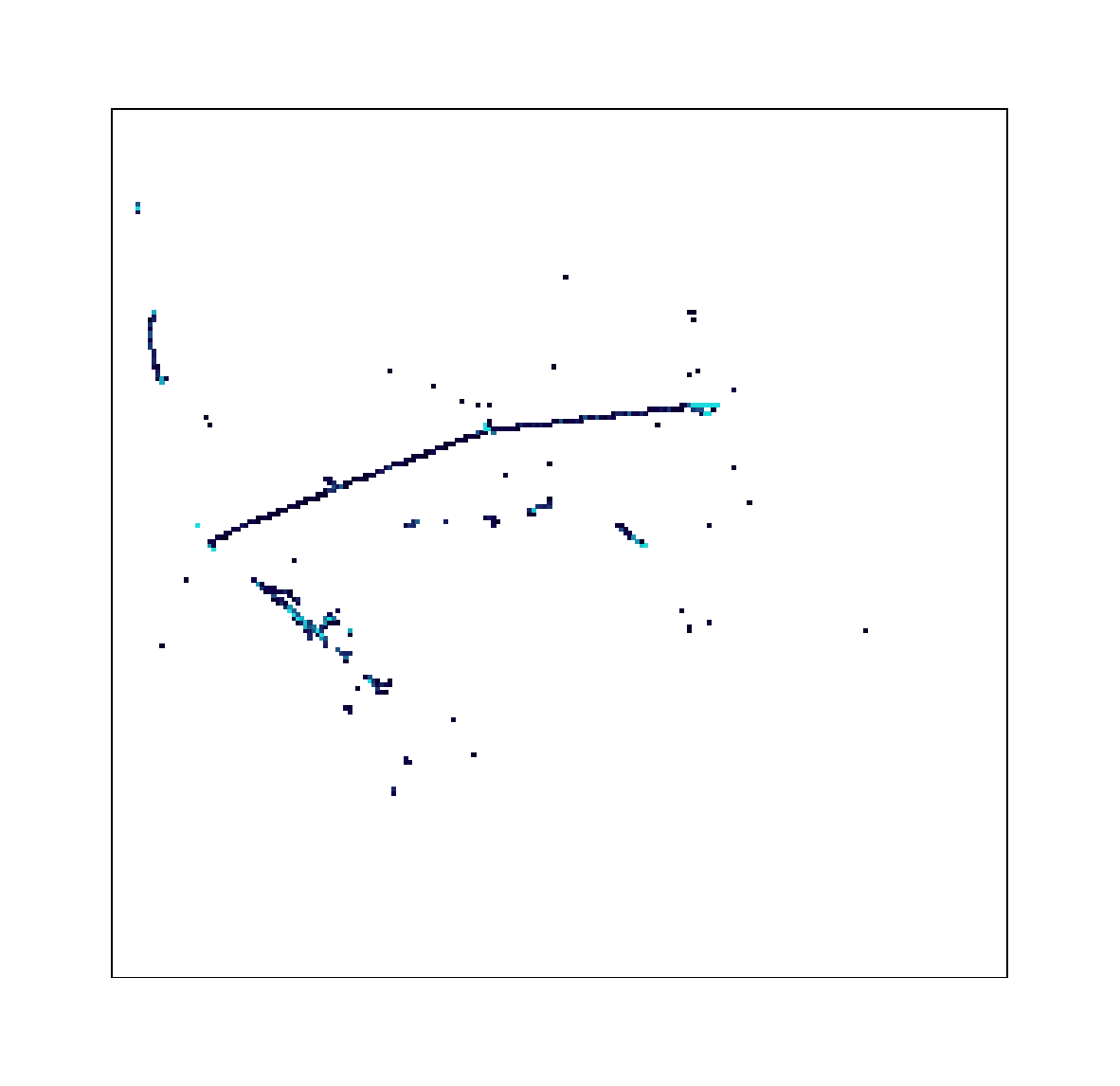}
    \caption{\label{fig:event_images}Examples of CC~$\nu_\mu$, CC~$\nu_e$ and NC interactions shown in the left, middle and right panels, respectively. The events are shown in the $w$ readout plane in the  $\left(w,x\right)$ parameter space and the colour scale, in arbitrary units, goes from dark blue (low charge) to cyan (high charge).}
\end{figure*}

Figure~\ref{fig:event_images} shows three neutrino interactions from the dataset: CC~$\nu_\mu$ (left), CC~$\nu_e$ (middle), and NC (right). The events are shown in the $w$ readout view only for clarity, but the images in the dataset have the three readout views stacked to mimic colour images with $\left(u,v,w\right)$ instead of $\left(r,g,b\right)$. The CC~$\nu_\mu$ interactions are typically characterised by a long muon track, CC~$\nu_e$ events by an electron-induced electromagnetic shower emanating from the interaction vertex, and NC events lack either of these features.

\section{Model Architectures}\label{sec:models}
\glspl{cnn} have historically improved image recognition performance by increasing network depth, width, and input resolution~\cite{he2015deep}, but this quickly becomes computationally heavy and harder to optimise \cite{zhang2024reducecomputationalcomplexityconvolutional}. To manage complexity while preserving performance, different network families take distinct approaches.

\subsection{Residual Neural Networks}

\glspl{resnet} \cite{he2015deepresiduallearningimage} use residual blocks with identity skip connections, which make it easier to train deeper networks. These skip connections help reduce vanishing-gradient problems and improve optimization stability. ResNets can achieve greater depth without the training degradation often seen in plain networks. ResNetV2 improves the original ResNet design by using pre-activation residual blocks, which support better gradient flow during training~\cite{he2016identity}. Because of their strong performance and reliability, ResNet-based architectures remain widely used as baseline models.

\subsection{Densely Connected Convolutional Networks}

\glspl{dense} use dense blocks, where each layer connects to every other layer in a feed-forward fashion. Within a dense block, for each layer, the feature maps of all preceding layers are used as inputs, and its own feature maps become inputs to all subsequent layers. This architecture alleviates vanishing gradients, strengthens feature propagation, encourages feature reuse, and substantially reduces parameter count. Dense block concepts allow for deep scaling and can also be employed outside DenseNets~\cite{huang2018densely}.

\subsection{Inception Networks}

InceptionNets \cite{szegedy2014goingdeeperconvolutions} circumvent depth scaling by employing multi-branch Inception blocks. These blocks process the same input through parallel convolutional paths and concatenate their outputs. Larger spatial filters are factorised into sequences of smaller ones to reduce computational cost. InceptionV2 introduces factorised convolutions and BatchNorm to reduce computation and improve training, while InceptionV3 further adds label smoothing, refined factorisation, and optimisation enhancements to increase depth and accuracy. Unlike ResNets and DenseNets, Inception families primarily widen rather than deepen networks~\cite{szegedy2016rethinking}.

\subsection{Efficient Networks}

EfficientNets \cite{tan2020efficientnetrethinkingmodelscaling} do not merely scale depth or width; instead, they use neural architecture search (NAS) to discover an optimal baseline network (B0) which is then compound-scaled—jointly adjusting depth, width, and resolution to balance accuracy and FLOPs. This produces state-of-the-art accuracy at drastically lower computational cost. EfficientNets use MBConv blocks (inverted bottlenecks with depthwise filtering and Squeeze-and-Excitation attention) with stage-wise downsampling. EfficientNetV2 replaces early MBConv blocks with Fused-MBConv (standard convolutions that merge expansion and filtering for faster early-stage training) and incorporates progressive resizing and regularisation. This unified scaling strategy yields compact, deployment-friendly models with strong accuracy–efficiency trade-offs~\cite{tan2020efficientnet}.

Across modern deep learning, the \glspl{cnn} examined in this study provide foundational building blocks underpinning many more complex architectures. Hybrids such as Inception-ResNet~\cite{szegedy2016inception} combine different block types to scale networks deeper and wider. In the particle-physics setting, this paper demonstrates the effectiveness of quantising these baseline models, suggesting that downstream, more elaborate \glspl{cnn} built from the same blocks can also be quantised and deployed without prohibitive accuracy loss. They were chosen as they are available as predefined networks in \texttt{keras}, are representative of architectures used in the field, and perform well on the neutrino dataset described in Section~\ref{sec:datasets}. 

\subsection{Training details} \label{sec:training}

The four \glspl{cnn} have been trained and subsequently fine-tuned using \gls{qat}. The resulting quantised models have been compiled and deployed on the Edge \gls{tpu}. \gls{ptq} was performed with the standard TensorFlow Lite converter using full-integer quantisation, with activation ranges calibrated on a representative subset of the training set and input and output tensors set to \texttt{uint8}.

\subsubsection{Data Handling}
The four convolutional models have been trained using TensorFlow~2.15 (tf.keras) on the neutrino dataset described in Section \ref{sec:datasets}. Input files are streamed directly from disk to prevent out-of-memory (OOM) issues, and no data augmentation is applied. The images are 3-channel PNGs (three views stacked), with inputs resized to $224\times224$, except for InceptionV3, which uses its canonical input size of $299\times299$. Input pixel values are scaled to $[0,1]$. All models were trained from scratch using the Adam optimiser \cite{Kingma_Ba_2017}. The learning rate starts at $1\times10^{-6}$, together with a \texttt{ReduceLROnPlateau} scheduler: if validation loss does not improve for 5 epochs, the learning rate is reduced by a factor of 0.1; if validation loss does not improve for 10 epochs, training stops early. A training set of 17,338 events and a test set of 5,000 events are used, as 
described in Section 4. A batch size of~64 is employed and epochs are capped generously (e.g.\ 5000 or 50000). EfficientNetV2B0 is the exception, using batch size~12.

\subsubsection{Checkpointing}
Per-epoch checkpoints are saved automatically, and the best checkpoint (based on validation loss) is continuously updated. A copy of checkpoints is also saved in HDF5 (.h5) format for convenience and compatibility.

\subsubsection{Logging and Evaluation}
The training script supports optional real-time logging with Weights \& Biases (wandb), including automatic run naming. After training, the script generates a test-set confusion matrix (using scikit-learn), a classification report, and the macro-averaged F1 score.

\subsubsection{Quantisation-Aware Training}
\gls{qat} was implemented with the TensorFlow Model Optimization Toolkit (TFMOT). Selective quantization is applied to \texttt{Conv2D}, \texttt{DepthwiseConv2D}, and \texttt{Dense} layers, while BatchNorm, ReLU, pooling, and padding layers are skipped. When available, a \texttt{float32} checkpoint is loaded for initialization, after which the quantized graph is fine-tuned so scale and zero-point parameters calibrate without large accuracy degradation. The learning-rate schedule used during \gls{qat} fine-tuning differs by architecture: EfficientNetV2B0 uses a per-step warmup over the first 10\% of total steps followed by cosine decay, while ResNet50V2, DenseNet169, and InceptionV3 use validation-driven \texttt{ReduceLROnPlateau} with factor~0.2, $\mathrm{min\_lr}=1\times10^{-7}$, and patience~5. The batch sizes used for the \gls{qat} fine-tuning stage only, which differ from those of the floating-point training described above, are set according to the available memory: ResNet50V2 uses batch~64; DenseNet169 uses approximately $\lfloor 64/1.4\rfloor\approx 45$; EfficientNetV2B0 uses batch~24; and InceptionV3 uses approximately $\lfloor 64/1.4\rfloor\approx 45$ at $299\times299$.

During fine-tuning, incomplete final batches are discarded so that all batches have the fixed input shape required by the quantised graph, and occasional images that cannot be decoded are skipped during evaluation. The best checkpoints are stored as weights only, and the final quantised models are stored together with their quantisation parameters so that they can be converted for deployment.

\subsubsection{Reproducibility, Environment, and Acknowledgment}
A fixed random seed is used for Python, NumPy, and TensorFlow to improve run-to-run reproducibility. GPU memory growth is enabled to reduce pre-allocation and avoid OOM. The environment includes TensorFlow~2.15 (tf.keras), Python~3.9.18 , CPU~AMD EPYC™ 7763, and GPU~Nvidia A100.

\section{Measurements}\label{sec:results}

After the models have been quantised using the \gls{qat} fine-tuning step (as described in Section \ref{sec:qat}) and \gls{ptq} (as described in Section \ref{sec:ptq}), they have been compiled for the Edge \gls{tpu} using the compiler version \texttt{16.0.384591198}. Subsequently, they have been deployed on the Edge \gls{tpu} using a Linux system and connecting the Edge \gls{tpu} via a USB-A to USB-C cable. The USB 3.0 interface provides a nominal bandwidth of 5\,Gbit/s, whereas each input image corresponds to approximately 150\,kB (268\,kB for InceptionV3), i.e.\ an input data rate of order 10\,MB/s at the latencies measured in this work. For comparison, the raw data rate of a DUNE far detector module is of order 1\,TB/s (384{,}000 channels sampled at 2\,MHz with 12-bit precision~\cite{DUNE:2020txw}); a single Edge \gls{tpu} could therefore not process the raw data stream, but could operate downstream of the data-reduction stages, on images of candidate events selected by lower-level triggers. The models were tasked with performing inference on the neutrino dataset described in Section \ref{sec:datasets} and identifying the neutrino interaction shown in each image. Results of the study are described below.

\subsection{Accuracy Degradation}

Table \ref{tab:accuracy} shows the results on balanced accuracy for the models following \gls{ptq} and \gls{qat} deployment pipelines. The response to quantisation differs strongly between models, as discussed in Section~\ref{sec:discussion}. Each model was trained once, with a fixed random seed, so the variation between training runs was not assessed. The statistical uncertainty on the balanced accuracy due to the finite size of the test set is approximately 0.5 percentage points; changes of order 1 percentage point, such as those observed for InceptionV3, are therefore not significant, whereas the larger degradations observed for the other models are.

\begin{table*}[t]
\centering
\resizebox{\textwidth}{!}{%
\begin{tabular}{lcccc}
\toprule
\multicolumn{5}{c}{\textbf{Post-Training Quantisation pipeline}} \\
\midrule
\textbf{Model Name} & \textbf{TF} & \textbf{PTQ} & \textbf{Edge TPU PTQ} & \\
\midrule
ResNet-50V2      & 79.26\% & 66.87\% (-12.39) & 61.47\% (-5.40) & \\
DenseNet-169     & 83.55\% & 70.07\% (-13.48) & 80.10\% (+10.03) & \\
EfficientNetV2B0 & 80.16\% & 33.33\% (-46.83) & 33.33\% (+0.00) & \\
InceptionV3      & 87.48\% & 87.14\% (-0.34)  & 87.07\% (-0.07) & \\
\midrule
\multicolumn{5}{c}{\textbf{Quantisation-Aware Training pipeline}} \\
\midrule
\textbf{Model Name} & \textbf{TF} & \textbf{QAT} & \textbf{QAT quantised} & \textbf{Edge TPU QAT} \\
\midrule
ResNet-50V2      & 79.26\% & 68.46\% (-10.80) & 67.52\% (-0.94) & 46.18\% (-21.34) \\
DenseNet-169     & 83.55\% & 83.40\% (-0.15)  & 81.16\% (-2.24) & 56.06\% (-25.10) \\
EfficientNetV2B0 & 80.16\% & 83.46\% (+3.30)  & 53.92\% (-29.54) & 33.40\% (-20.52) \\
InceptionV3      & 87.48\% & 88.62\% (+1.14)  & 88.10\% (-0.52) & 88.18\% (+0.08) \\
\bottomrule
\end{tabular}%
}
\caption{Balanced accuracy for the two deployment pipelines considered in this work. Values in parentheses indicate the change in balanced accuracy with respect to the previous stage (percentage points).}
\label{tab:accuracy}
\end{table*}

\subsection{Processing Time and Power Usage}

Latency and energy were measured on the \gls{cpu} and \gls{gpu} platforms listed in Table~\ref{tab:devices} at batch sizes of 1, 16 and 128, and on the Edge \gls{tpu}, which supports only a batch size of one~\cite{CoralModelReqs}. On the \gls{cpu}, batch size 1 was measured with 1, 2, 4, 8 and 16 threads, and batch sizes 16 and 128 with 16 threads. On the \gls{cpu} and \glspl{gpu}, inference was executed directly in TensorFlow, without deployment-oriented runtimes such as ONNX Runtime or NVIDIA Triton, using TensorFlow~2.19 (Keras~3) for ResNet-50V2, DenseNet-169 and EfficientNetV2B0, and TensorFlow~2.15 (Keras~2) for InceptionV3 owing to a compatibility issue. For each configuration, latency was measured over 300 timed inference calls, in three repetitions of 100 after 20 untimed warm-up calls, cycling over 100 to 128 different events; it is reported as the time per call divided by the batch size, using the mean on the A100 and the median on the \gls{cpu} and RTX 4060, for which the mean agrees with the median within 5\%. On the Edge \gls{tpu}, inference was executed with PyCoral, and latency is the mean over one timed inference per event of the test set, after 10 untimed warm-up inferences. The latency includes the transfer of the input from host memory to the device, inference and the return of the output to host memory; on the Edge \gls{tpu}, this comprises setting the input tensor, inference including USB communication, and retrieving the output. Image loading, preprocessing and output post-processing are excluded.

Energy was measured over a separate 10\,s window of continuous inference per configuration, and energy per event is the energy consumed in the window divided by the number of events processed. On the A100, energy was read from the cumulative energy counter of the NVIDIA Management Library (NVML), polled every 0.1\,s; the A100 was used exclusively during the measurements. On the RTX 4060, energy was obtained from NVML power readings sampled every 0.1\,s, each being a 1\,s average, which agree with an independent \texttt{nvidia-smi} log within 2\% (6\% for InceptionV3 at batch size 128). On the \gls{cpu}, energy was read from the Running Average Power Limit (RAPL) package energy counter every 0.2\,s. These measurements include the idle power of the device, which was not subtracted (approximately 16\,W for the \gls{cpu}), and exclude the rest of the host system. Each configuration was measured once. The highest mean power measured on the A100 was 340\,W (InceptionV3, batch size 128), corresponding to 85\% of its 400\,W \gls{tdp}, at a mean sampled utilisation of 94--95\%. The power of the Edge \gls{tpu} was not measured; its energy per event is estimated as its nominal power of 2\,W multiplied by the latency, and likewise excludes the host system.

\begin{table*}[t]
\centering
\small
\begin{tabular*}{\textwidth}{@{\extracolsep{\fill}}llcccccccc}
\toprule
 & & \multicolumn{2}{c}{\makebox[0pt]{\textbf{ResNet-50V2}}} & \multicolumn{2}{c}{\makebox[0pt]{\textbf{DenseNet-169}}} & \multicolumn{2}{c}{\makebox[0pt]{\textbf{EfficientNetV2B0}}} & \multicolumn{2}{c}{\makebox[0pt]{\textbf{InceptionV3}}} \\
\cmidrule(lr){3-4}\cmidrule(lr){5-6}\cmidrule(lr){7-8}\cmidrule(lr){9-10}
\textbf{Device} & \textbf{Batch} & $t$ [ms] & $E$ [mJ] & $t$ [ms] & $E$ [mJ] & $t$ [ms] & $E$ [mJ] & $t$ [ms] & $E$ [mJ] \\
\midrule
\multirow{3}{*}{\begin{tabular}[c]{@{}l@{}}Intel i7-13650HX\\(16 threads)\end{tabular}} & 1   & 47.40 & 2692 & 83.93 & 4876 & 34.46 & 2078 & 56.67 & 3539 \\
                                                & 16  & 17.80 & 970  & 29.00 & 1686 & 8.11  & 495  & 22.80 & 1225 \\
                                                & 128 & 17.22 & 928  & 31.58 & 1715 & 7.97  & 439  & 20.27 & 1070 \\
\midrule
\multirow{3}{*}{\begin{tabular}[c]{@{}l@{}}NVIDIA RTX 4060\\Laptop\end{tabular}} & 1   & 7.19  & 401  & 19.89 & 622  & 7.73  & 214  & 8.45  & 483 \\
                                                & 16  & 2.04  & 182  & 3.29  & 230  & 1.11  & 67   & 2.03  & 217 \\
                                                & 128 & 2.10  & 201  & 3.42  & 296  & 1.09  & 90   & 1.99  & 216 \\
\midrule
\multirow{3}{*}{\begin{tabular}[c]{@{}l@{}}NVIDIA A100\\SXM4\end{tabular}} & 1   & 6.00  & 856  & 15.61 & 1814 & 6.33  & 728  & 8.11  & 1169 \\
                                                & 16  & 0.91  & 244  & 1.61  & 340  & 0.58  & 108  & 1.02  & 282 \\
                                                & 128 & 0.67  & 227  & 0.93  & 282  & 0.30  & 89   & 0.66  & 226 \\
\midrule
Edge TPU (PTQ)                                  & 1   & 39.03 & 78$^{*}$ & 20.59 & 41$^{*}$ & 12.32 & 25$^{*}$ & 39.85 & 80$^{*}$ \\
Edge TPU (QAT)                                  & 1   & 41.21 & 82$^{*}$ & 21.99 & 44$^{*}$ & 13.91 & 28$^{*}$ & 40.08 & 80$^{*}$ \\
\bottomrule
\end{tabular*}
\caption{Latency $t$ and energy $E$ per event as a function of batch size. Latency is the mean on the A100 and the Edge \gls{tpu} and the median on the \gls{cpu} and RTX 4060. \gls{cpu} and \gls{gpu} energies are measured. $^{*}$Estimated as nominal power (2\,W) $\times$ latency.}
\label{tab:latency_energy}
\end{table*}

\begin{table*}[t]
\centering
\small
\begin{tabular*}{\textwidth}{@{\extracolsep{\fill}}lrrrrrrrr}
\toprule
 & \multicolumn{2}{c}{\textbf{ResNet-50V2}} & \multicolumn{2}{c}{\textbf{DenseNet-169}} & \multicolumn{2}{c}{\textbf{EfficientNetV2B0}} & \multicolumn{2}{c}{\textbf{InceptionV3}} \\
\cmidrule(lr){2-3}\cmidrule(lr){4-5}\cmidrule(lr){6-7}\cmidrule(lr){8-9}
\textbf{Threads} & $t$ [ms] & $E$ [mJ] & $t$ [ms] & $E$ [mJ] & $t$ [ms] & $E$ [mJ] & $t$ [ms] & $E$ [mJ] \\
\midrule
1  & 71.59 & 2408 & 89.53 & 2876 & 20.94 & 786  & 99.68 & 3175 \\
2  & 45.18 & 1983 & 67.31 & 2874 & 22.14 & 948  & 64.32 & 2751 \\
4  & 38.74 & 1958 & 65.77 & 3218 & 24.77 & 1202 & 51.03 & 2703 \\
8  & 45.74 & 2442 & 78.83 & 4046 & 29.28 & 1438 & 59.92 & 3525 \\
16 & 47.40 & 2692 & 83.93 & 4876 & 34.46 & 2078 & 56.67 & 3539 \\
\bottomrule
\end{tabular*}
\caption{Latency $t$ and energy $E$ per event on the Intel i7-13650HX at batch size 1 as a function of the number of threads.}
\label{tab:cpu_threads}
\end{table*}

Table~\ref{tab:latency_energy} shows that at batch size 1, the case relevant for triggering, the A100 is the fastest device (6.0--15.6\,ms), with the RTX 4060 within a factor of 1.0--1.3 of it. The Edge \gls{tpu} is 1.3--6.5 times slower than the A100, and between equal to and 3.2 times faster than the \gls{cpu} in its fastest thread configuration (Table~\ref{tab:cpu_threads}). In terms of energy per event at batch size 1, the Edge \gls{tpu} requires 5--15 times less than the RTX 4060, 11--44 times less than the A100 and 25--70 times less than the \gls{cpu} in its lowest-energy thread configuration. Batching reduces the energy per event of the \glspl{gpu} substantially: at batch size 128, the Edge \gls{tpu} advantage is reduced to a factor of 2.6--7.2 with respect to both \glspl{gpu}, which reach similar energies per event. On the \gls{cpu}, latency at batch size 1 is lowest with 4 threads (1 thread for EfficientNetV2B0), and energy per event is lowest with 1--4 threads, depending on the model. The ranges above span all four models. For InceptionV3, the only model with no significant accuracy loss along either pipeline and therefore taken as the reference architecture, the Edge \gls{tpu} at batch size 1 is 1.3 times faster than the \gls{cpu} and 4.7--4.9 times slower than the \glspl{gpu}, and requires 6, 15 and 34 times less energy per event than the RTX 4060, the A100 and the \gls{cpu}, respectively; at batch size 128 this advantage is reduced to a factor of 2.7--2.8.

On the Edge \gls{tpu}, the \gls{ptq} models are consistently faster than the corresponding \gls{qat} models, by 0.2--2.2\,ms. Since both are compiled from the same architectures with the same compiler, the difference arises from differences between the quantised graphs produced by the two pipelines; as the compiler is proprietary, these could not be examined in detail.

\begin{figure*}[t]
    \centering
    \includegraphics[width=\textwidth]{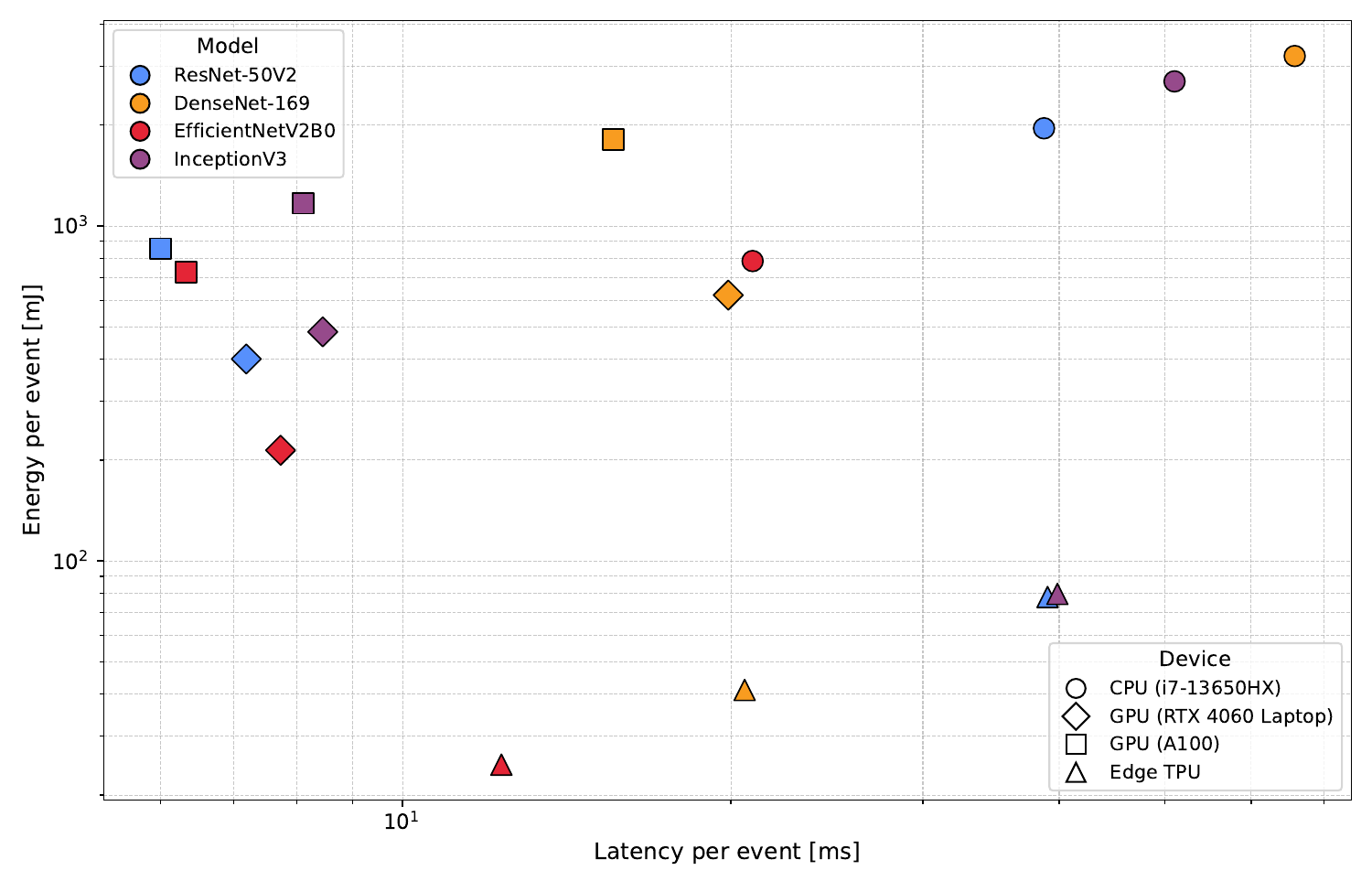}
\caption{Energy per event as a function of latency per event at batch size 1 for the four models, with each device in its fastest configuration: the \gls{cpu} with the number of threads giving the lowest latency for each model, and the Edge \gls{tpu} with the \gls{ptq} models. Edge \gls{tpu} energies are estimated from its nominal power. Colour palette has been chosen to be distinguishable for people with colour-vision deficiencies \cite{petroff2024accessiblecolorsequencesdata}.}
    \label{fig:energy_inference}
\end{figure*}

\subsection{Discussion}\label{sec:discussion}

As shown in Table \ref{tab:accuracy}, InceptionV3 is the best-performing model, with no significant accuracy loss along either pipeline. For the other models, a large part of the degradation already occurs at the quantisation step, before deployment on the Edge \gls{tpu}. EfficientNetV2B0 is the most affected: after \gls{ptq} its balanced accuracy drops to 33.33\%, the chance level for three classes, which corresponds to the model predicting a single class. It also shows the largest drop when the \gls{qat} model is converted to integer arithmetic ($-29.54$ percentage points). Its architecture contains elements known to be sensitive to quantisation, namely depthwise convolutions, the Swish activation and squeeze-and-excitation blocks, which were removed in EfficientNet-Lite, the variant designed for integer accelerators such as the Edge \gls{tpu}~\cite{EfficientNetLite}. Techniques designed to ease quantisation, such as quantisation-friendly activation functions, range clipping or cross-layer equalisation~\cite{Nagel2021WhitePaper}, were not applied in this work; they could reduce the degradation observed for ResNet-50V2, DenseNet-169 and EfficientNetV2B0, and will be explored in future work.

Further variations occur at deployment. Since the compiled model performs essentially the same integer operations as the quantised TensorFlow Lite model, only small differences would be expected between the two. The large drops observed in the \gls{qat} pipeline for ResNet-50V2, DenseNet-169 and EfficientNetV2B0, and the improvement of DenseNet-169 in the \gls{ptq} pipeline, are therefore not explained by quantisation itself. The compiler used for this work is proprietary, which made it impossible to determine their origin.

At batch size 1, across the four models, the Edge \gls{tpu} is comparable to or faster than the \gls{cpu} and slower than both \glspl{gpu}, and requires between 5 and 70 times less energy per event than the other devices (6--34 times for InceptionV3). With batching, the \glspl{gpu} reduce this gap to a factor of 2.6--7.2. Two caveats apply to the energy comparison: none of the measurements includes the host system, and the \gls{cpu} and \gls{gpu} figures include the idle power of the device, whereas the Edge \gls{tpu} figure is an estimate from its nominal power; and the \gls{cpu} and \gls{gpu} baselines run \texttt{float32} models, whereas the Edge \gls{tpu} runs \texttt{uint8} models. In the energy--latency plane (Figure~\ref{fig:energy_inference}), the devices are clearly separated: the \glspl{gpu} are the fastest option, the Edge \gls{tpu} the least energy-consuming, and the \gls{cpu} the least favourable in both respects.

The relevant comparison depends on the use case. For live triggering, events must be processed individually as they arrive, and the comparison at batch size 1 applies: here the Edge \gls{tpu} offers the lowest energy per event and the lowest power, at a latency comparable to or lower than that of the \gls{cpu}. For offline analysis of large datasets, events can be batched, and the relevant comparison is with each device in its best configuration: at batch size 128, the \glspl{gpu} process events 6--60 times faster than the Edge \gls{tpu} across the four models (20--60 times for InceptionV3), while the Edge \gls{tpu} still requires 2.6--7.2 times less energy per event (2.7--2.8 times for InceptionV3). For offline analysis, the Edge \gls{tpu} is therefore advantageous where energy consumption, rather than processing time, is the main constraint.

\section{Conclusion}\label{sec:conclusion}

This work shows that it is possible to deploy large \gls{cnn} models to perform neutrino interaction recognition on the Edge \gls{tpu}, although the accuracy degradation introduced by quantisation and deployment depends strongly on the model: InceptionV3 shows no significant loss along either pipeline, whereas the other models are significantly degraded. The benchmark includes two classes of \gls{gpu}: a consumer RTX 4060, representative of the hardware available on a scientist's workstation, and a state-of-the-art data-centre A100. At batch size 1, the case relevant for triggering, and taking InceptionV3 as the reference architecture, the Edge \gls{tpu} is 1.3 times faster than the \gls{cpu} and 4.7--4.9 times slower than both \glspl{gpu}, and requires 6 to 34 times less energy per event than the other devices considered (5 to 70 times across all four models); with batching, the \glspl{gpu} reduce this gap to a factor of 2.7--2.8 (2.6--7.2 across all models). The Edge \gls{tpu} is also approximately four times cheaper than the RTX 4060, whose latency at batch size 1 is within a factor of 1.3 of that of the A100, and more than two orders of magnitude cheaper than the A100.

The pipeline to quantise and deploy models for the Edge \gls{tpu} is relatively straightforward, and the device can be connected to any computer via a USB-A to USB-C cable. However, the accuracy of the deployed model must be validated for each architecture, since quantisation and compilation significantly degrade several of the models studied here. For models that retain their accuracy after quantisation, Edge \glspl{tpu} could offer a low-power alternative to \glspl{gpu} in live triggering, where events are processed individually, and in offline analysis where energy consumption rather than processing time is the main constraint. Their main advantage is the low power per device, approximately 2\,W compared with tens to hundreds of watts for \glspl{cpu} and \glspl{gpu}, which eases deployment where power and cooling are constrained, such as in underground detector halls, with an additional benefit in terms of environmental impact.

This work serves as a proof-of-concept that edge \gls{ai} devices can be used to perform low-power inference in neutrino, particle physics, and other science experiments. They could be valuable tools for live triggering in \glspl{lartpc}, operating on candidate events selected by lower-level triggers. One example could be identifying supernova neutrino signatures \cite{Scholberg2018SupernovaMassOrdering,DUNE2025SupernovaPointing,cuesta2024supernovasolarneutrinosearches} and rare decays \cite{domingo2024novelprotondecaysignature} in experiments such as \gls{dune}.

\backmatter

\bibliography{sn-bibliography}

@article{Chappell:2022yxd,
    author = "Chappell, Andrew and Whitehead, Leigh H.",
    title = "{Application of transfer learning to neutrino interaction classification}",
    doi = "10.1140/epjc/s10052-022-11066-6",
    journal = "Eur. Phys. J. C",
    volume = "82",
    number = "12",
    pages = "1099",
    year = "2022"
}

@misc{seshadri2022evaluation,
      title={An Evaluation of Edge TPU Accelerators for Convolutional Neural Networks}, 
      author={Kiran Seshadri and Berkin Akin and James Laudon and Ravi Narayanaswami and Amir Yazdanbakhsh},
      year={2022},
      eprint={2102.10423},
      archivePrefix={arXiv},
      primaryClass={cs.LG}
}

@article{Yazdanbakhsh:2021,
    author = "Yazdanbakhsh, Amir and Seshadri, Kiran and Akin, Berkin and Laudon, James and Narayanaswami, Ravi",
    title = "{An Evaluation of Edge TPU Accelerators for Convolutional Neural Networks}",
    eprint = "2102.10423",
    archivePrefix = "arXiv",
    primaryClass = "cs.AR",
    year = "2021"
}

@article{Sun:2021,
    author = "Sun, Y. and Kist, A. M.",
    title = "{Deep Learning on Edge TPUs}",
    eprint = "2108.13732",
    archivePrefix = "arXiv",
    primaryClass = "cs.LG",
    year = "2021"
}

@article{Jacob:2017,
    author = "Jacob, Benoit and Kligys, Skirmantas and Chen, Bo and Zhu, Menglong and Tang, Matthew and Howard, Andrew and Adam, Hartwig and Kalenichenko, Dmitry",
    title = "{Quantization and Training of Neural Networks for Efficient Integer-Arithmetic-Only Inference}",
    eprint = "1712.05877",
    archivePrefix = "arXiv",
    primaryClass = "cs.LG",
    year = "2017"
}

@misc{Alam:2015nkk,
  doi = {10.48550/ARXIV.1512.06882},
  url = {https://arxiv.org/abs/1512.06882},
  author = {Alam,  M. and others},
  title = {GENIE Production Release 2.10.0},
  publisher = {arXiv},
  year = {2015},
  copyright = {arXiv.org perpetual,  non-exclusive license}
}

@article{Agostinelli:2002hh,
    author         = "Agostinelli, S. and others",
    title          = "{GEANT4: A Simulation toolkit}",
    collaboration  = "GEANT4",
    journal        = "Nucl.\ Instrum.\ Meth.\ A",
    volume         = "506",
    year           = "2003",
    pages          = "250-303",
    doi            = "10.1016/S0168-9002(03)01368-8",
    reportNumber   = "SLAC-PUB-9350, FERMILAB-PUB-03-339",
    SLACcitation   = "%%CITATION = NUIMA,A506,250;%%"
}

@article{DUNE:2020jqi,
    author = "Abi, B. and others",
    collaboration = "DUNE",
    title = "{Long-baseline neutrino oscillation physics potential of the DUNE experiment}",
    reportNumber = "FERMILAB-PUB-20-251-E-LBNF-ND-PIP2-SCD, PUB-20-251-E-LBNF-ND-PIP2-SCD",
    doi = "10.1140/epjc/s10052-020-08456-z",
    journal = "Eur. Phys. J. C",
    volume = "80",
    number = "10",
    pages = "978",
    year = "2020"
}

@article{DUNE:2020txw,
    author = "Abi, Babak and others",
    collaboration = "DUNE",
    title = "{Deep Underground Neutrino Experiment (DUNE), Far Detector Technical Design Report, Volume IV: Far Detector Single-phase Technology}",
    reportNumber = "FERMILAB-PUB-20-027-ND, FERMILAB-DESIGN-2020-04",
    doi = "10.1088/1748-0221/15/08/T08010",
    journal = "JINST",
    volume = "15",
    number = "08",
    pages = "T08010",
    year = "2020"
}

@article{Rubbia:1977zz,
    author = "Rubbia, C.",
    title = "{The Liquid Argon Time Projection Chamber: A New Concept for Neutrino Detectors}",
    reportNumber = "CERN-EP-INT-77-08, CERN-EP-77-08",
    month = "5",
    year = "1977"
}

@article{Cavanna:2014iqa,
    author = "Cavanna, F. and Kordosky, M. and Raaf, J. and Rebel, B.",
    collaboration = "LArIAT",
    title = "{LArIAT: Liquid Argon In A Testbeam}",
    eprint = "1406.5560",
    archivePrefix = "arXiv",
    reportNumber = "FERMILAB-PUB-14-268-E",
    month = "6",
    year = "2014"
}

@misc{guenette2011argoneutexperiment,
      title={The ArgoNeuT experiment}, 
      author={Roxanne Guenette},
      year={2011},
      eprint={1110.0443},
      archivePrefix={arXiv},
      primaryClass={physics.ins-det},
      url={https://arxiv.org/abs/1110.0443}, 
}

@misc{icaruscollaboration2023icarusfermilabshortbaselineneutrino,
      title={ICARUS at the Fermilab Short-Baseline Neutrino Program -- Initial Operation}, 
      author={ICARUS Collaboration},
      year={2023},
      eprint={2301.08634},
      archivePrefix={arXiv},
      primaryClass={hep-ex},
      url={https://arxiv.org/abs/2301.08634}, 
}

@article{MicroBooNE:2016pwy,
    author = "Acciarri, R. and others",
    collaboration = "MicroBooNE",
    title = "{Design and Construction of the MicroBooNE Detector}",
    eprint = "1612.05824",
    archivePrefix = "arXiv",
    primaryClass = "physics.ins-det",
    reportNumber = "FERMILAB-PUB-16-613-ND",
    doi = "10.1088/1748-0221/12/02/P02017",
    journal = "JINST",
    volume = "12",
    number = "02",
    pages = "P02017",
    year = "2017"
}

@article{nutelescopetpu,
author = {Jin, Miaochen and Hu, Yushi and Argüelles, C.A.},
year = {2024},
month = {06},
pages = {},
title = {Two Watts is all you need: enabling in-detector real-time machine learning for neutrino telescopes via edge computing},
volume = {2024},
journal = {Journal of Cosmology and Astroparticle Physics},
doi = {10.1088/1475-7516/2024/06/026}
}

@misc{Kingma_Ba_2017, 
title={Adam: A method for stochastic optimization}, 
url={https://arxiv.org/abs/1412.6980}, 
journal={arXiv.org}, 
author={Kingma, Diederik P. and Ba, Jimmy}, 
year={2017}, 
month={Jan}}

@article{he2015deep,
  title={Deep Residual Learning for Image Recognition},
  author={He, Kaiming and Zhang, Xiangyu and Ren, Shaoqing and Sun, Jian},
  journal={arXiv preprint arXiv:1512.03385},
  year={2015}
}

@inproceedings{he2016identity,
  author    = {He, Kaiming and Zhang, Xiangyu and Ren, Shaoqing and Sun, Jian},
  title     = {Identity Mappings in Deep Residual Networks},
  booktitle = {Computer Vision -- ECCV 2016},
  series    = {Lecture Notes in Computer Science},
  volume    = {9908},
  pages     = {630--645},
  year      = {2016},
  publisher = {Springer},
  address   = {Cham},
  doi       = {10.1007/978-3-319-46493-0_38}
}

@article{szegedy2016inception,
  title={Inception-v4, Inception-ResNet and the Impact of Residual Connections on Learning},
  author={Szegedy, Christian and Ioffe, Sergey and Vanhoucke, Vincent and Alemi, Alex},
  journal={arXiv preprint arXiv:1602.07261},
  year={2016}
}

@article{huang2018densely,
  title={Densely Connected Convolutional Networks},
  author={Huang, Gao and Liu, Zhuang and Van Der Maaten, Laurens and Weinberger, Kilian Q},
  journal={arXiv preprint arXiv:1608.06993},
  year={2018}
}

@inproceedings{szegedy2016rethinking,
  title={Rethinking the Inception Architecture for Computer Vision},
  author={Szegedy, Christian and Vanhoucke, Vincent and Ioffe, Sergey and Shlens, Jon and Wojna, Zbigniew},
  booktitle={Proceedings of the IEEE Conference on Computer Vision and Pattern Recognition (CVPR)},
  pages={2818--2826},
  year={2016}
}

@article{tan2020efficientnet,
  title={EfficientNet: Rethinking Model Scaling for Convolutional Neural Networks},
  author={Tan, Mingxing and Le, Quoc},
  journal={arXiv preprint arXiv:1905.11946},
  year={2020}
}

@misc{zhang2023posttrainingquantizationneuralnetworks,
      title={Post-training Quantization for Neural Networks with Provable Guarantees}, 
      author={Jinjie Zhang and Yixuan Zhou and Rayan Saab},
      year={2023},
      eprint={2201.11113},
      archivePrefix={arXiv},
      url={https://arxiv.org/abs/2201.11113}, 
}

@misc{zhang2024reducecomputationalcomplexityconvolutional,
      title={Reduce Computational Complexity for Convolutional Layers by Skipping Zeros}, 
      author={Zhiyi Zhang and Pengfei Zhang and Zhuopin Xu and Qi Wang},
      year={2024},
      eprint={2306.15951},
      archivePrefix={arXiv},
      primaryClass={cs.LG},
      url={https://arxiv.org/abs/2306.15951}, 
}

@misc{he2015deepresiduallearningimage,
      title={Deep Residual Learning for Image Recognition}, 
      author={Kaiming He and Xiangyu Zhang and Shaoqing Ren and Jian Sun},
      year={2015},
      eprint={1512.03385},
      archivePrefix={arXiv},
      primaryClass={cs.CV},
      url={https://arxiv.org/abs/1512.03385}, 
}

@misc{szegedy2014goingdeeperconvolutions,
      title={Going Deeper with Convolutions}, 
      author={Christian Szegedy and Wei Liu and Yangqing Jia and Pierre Sermanet and Scott Reed and Dragomir Anguelov and Dumitru Erhan and Vincent Vanhoucke and Andrew Rabinovich},
      year={2014},
      eprint={1409.4842},
      archivePrefix={arXiv},
      primaryClass={cs.CV},
      url={https://arxiv.org/abs/1409.4842}, 
}

@misc{tan2020efficientnetrethinkingmodelscaling,
      title={EfficientNet: Rethinking Model Scaling for Convolutional Neural Networks}, 
      author={Mingxing Tan and Quoc V. Le},
      year={2020},
      eprint={1905.11946},
      archivePrefix={arXiv},
      primaryClass={cs.LG},
      url={https://arxiv.org/abs/1905.11946}, 
}

@phdthesis{Vergani:2024syg,
    author = "Vergani, Stefano",
    title = "{Using Cutting Edge Software and Techniques to Model and Measure Experimental Neutrino Data}",
    doi = "10.17863/CAM.113334",
    school = "Cambridge U.",
    year = "2024"
}

@article{ALNAFRAH2025126813,
title = {The Two Tales of AI: A Global assessment of the environmental impacts of artificial intelligence from a multidimensional policy perspective},
journal = {Journal of Environmental Management},
volume = {392},
pages = {126813},
year = {2025},
issn = {0301-4797},
doi = {https://doi.org/10.1016/j.jenvman.2025.126813},
url = {https://www.sciencedirect.com/science/article/pii/S0301479725027896},
author = {Ibrahim Alnafrah}
}

@article{Wang2024AIenvironment,
  author       = {Qiang Wang and Yuanfan Li and Rongrong Li},
  title        = {Ecological footprints, carbon emissions, and energy transitions: the impact of artificial intelligence (AI)},
  journal      = {Humanities and Social Sciences Communications},
  year         = {2024},
  volume       = {11},
  pages        = {1043},
  doi          = {10.1057/s41599-024-03520-5},
  url          = {https://doi.org/10.1057/s41599-024-03520-5}
}

@misc{elsworth2025measuringenvironmentalimpactdelivering,
      title={Measuring the environmental impact of delivering AI at Google Scale}, 
      author={Cooper Elsworth and Keguo Huang and David Patterson and Ian Schneider and Robert Sedivy and Savannah Goodman and Ben Townsend and Parthasarathy Ranganathan and Jeff Dean and Amin Vahdat and Ben Gomes and James Manyika},
      year={2025},
      eprint={2508.15734},
      archivePrefix={arXiv},
      primaryClass={cs.AI},
      url={https://arxiv.org/abs/2508.15734}, 
}

@misc{falk2025carboncradletograveenvironmentalimpacts,
      title={More than Carbon: Cradle-to-Grave environmental impacts of GenAI training on the Nvidia A100 GPU}, 
      author={Sophia Falk and others},
      year={2025},
      eprint={2509.00093},
      archivePrefix={arXiv},
      primaryClass={cs.CY},
      url={https://arxiv.org/abs/2509.00093}, 
}

@article{Nik2025DecodingEnergy,
  author       = {Alireza Nik and Michael A. Riegler and P{\aa}l Halvorsen},
  title        = {Impact of decoding strategies on GPU energy usage in large language model text generation},
  journal      = {Scientific Reports},
  year         = {2025},
  volume       = {15},
  pages        = {31896},
  doi          = {10.1038/s41598-025-31896-0},
  url          = {https://doi.org/10.1038/s41598-025-31896-0}
}

@article{Schwartz2021Modern,
	author = {Schwartz, Matthew D.},
	journal = {Harvard Data Science Review},
	number = {2},
	year = {2021},
	month = {may 13},
	note = {https://hdsr.mitpress.mit.edu/pub/xqle7lat},
	publisher = {The MIT Press},
	title = {Modern {Machine} {Learning} and {Particle} {Physics}},
	volume = {3},
}

@article{BhattacherjeeMukherjee2024,
  author    = {Bhattacherjee, Biplob and Mukherjee, Swagata},
  title     = {Modern machine learning and particle physics: an in-depth review},
  journal   = {The European Physical Journal Special Topics},
  year      = {2024},
  doi       = {10.1140/epjs/s11734-024-01364-3},
  publisher = {Springer},
  note      = {Editorial}
}

@misc{liu2020deeplearningbasedkinematicreconstructiondune,
      title={Deep-Learning-Based Kinematic Reconstruction for DUNE}, 
      author={Junze Liu and others},
      year={2020},
      eprint={2012.06181},
      archivePrefix={arXiv},
      primaryClass={physics.ins-det},
      url={https://arxiv.org/abs/2012.06181}, 
}

@article{Albrecht2025TriggerLHC,
  author  = {Albrecht, J. and others},
  title   = {Summary of the trigger systems of the Large Hadron Collider experiments ALICE, ATLAS, CMS and LHCb},
  journal = {Journal of Physics G: Nuclear and Particle Physics},
  volume  = {52},
  number  = {3},
  pages   = {030501},
  year    = {2025},
  doi     = {10.1088/1361-6471/adaadc}
}

@article{Abratenko2026MicroBooNE,
  author  = {Abratenko, P. and others},
  title   = {Demonstration and performance of online data selection for liquid argon time projection chambers using MicroBooNE data},
  journal = {Journal of Instrumentation},
  volume  = {21},
  number  = {09},
  pages   = {P09044},
  year    = {2026},
  doi     = {10.1088/1748-0221/21/09/P09044}
}

@misc{ammendola2016gpubasedrealtimetriggeringna62,
      title={GPU-based Real-time Triggering in the NA62 Experiment}, 
      author={R. Ammendola and A. Biagioni and P. Cretaro and S. Di Lorenzo and R. Fantechi and M. Fiorini and O. Frezza and G. Lamanna and F. Lo Cicero and A. Lonardo and M. Martinelli and I. Neri and P. S. Paolucci and E. Pastorelli and R. Piandani and L. Pontisso and D. Rossetti and F. Simula and M. Sozzi and P. Vicini},
      year={2016},
      eprint={1606.04099},
      archivePrefix={arXiv},
      primaryClass={physics.ins-det},
      url={https://arxiv.org/abs/1606.04099}, 
}

@article{Duarte2018hls4ml,
  author        = {Duarte, Javier and Han, Song and Harris, Philip and Jindariani, Sergo and Kreinar, Edward and Kreis, Benjamin and Ngadiuba, Jennifer and Pierini, Maurizio and Rivera, Ryan and Tran, Nhan and Wu, Zhenbin},
  title         = {Fast inference of deep neural networks in {FPGAs} for particle physics},
  journal       = {JINST},
  volume        = {13},
  number        = {07},
  pages         = {P07027},
  year          = {2018},
  doi           = {10.1088/1748-0221/13/07/P07027},
  eprint        = {1804.06913},
  archivePrefix = {arXiv},
  primaryClass  = {physics.ins-det}
}

@ARTICLE{10.3389/frai.2022.855184,
    
AUTHOR={Jwa, Yeon-jae  and Di Guglielmo, Giuseppe  and Arnold, Lukas  and Carloni, Luca  and Karagiorgi, Georgia },
           
TITLE={Real-Time Inference With 2D Convolutional Neural Networks on Field Programmable Gate Arrays for High-Rate Particle Imaging Detectors},
          
JOURNAL={Frontiers in Artificial Intelligence},
          
VOLUME={Volume 5 - 2022},
  
YEAR={2022},
  
URL={https://www.frontiersin.org/journals/artificial-intelligence/articles/10.3389/frai.2022.855184},
  
DOI={10.3389/frai.2022.855184},
  
ISSN={2624-8212}}

@misc{wang2019benchmarkingtpugpucpu,
      title={Benchmarking TPU, GPU, and CPU Platforms for Deep Learning}, 
      author={Yu Emma Wang and Gu-Yeon Wei and David Brooks},
      year={2019},
      eprint={1907.10701},
      archivePrefix={arXiv},
      primaryClass={cs.LG},
      url={https://arxiv.org/abs/1907.10701}, 
}

@incollection{Godoy_2025,
  author    = {Godoy, William F. and others},
  title     = {Characterizing GPU Energy Usage in Exascale-Ready Portable Science Applications},
  booktitle = {High Performance Computing},
  publisher = {Springer Nature Switzerland},
  address   = {Cham},
  year      = {2025},
  pages     = {177--190},
  doi       = {10.1007/978-3-032-07612-0_14}
}

@misc{falk2025flopsfootprintsresourcecost,
      title={From FLOPs to Footprints: The Resource Cost of Artificial Intelligence}, 
      author={Sophia Falk and Nicholas Kluge Corrêa and Sasha Luccioni and Lisa Biber-Freudenberger and Aimee van Wynsberghe},
      year={2025},
      eprint={2512.04142},
      archivePrefix={arXiv},
      primaryClass={cs.CY},
      url={https://arxiv.org/abs/2512.04142}, 
}

@misc{savard2023optimizinghighthroughputinference,
      title={Optimizing High Throughput Inference on Graph Neural Networks at Shared Computing Facilities with the NVIDIA Triton Inference Server}, 
      author={Claire Savard and others},
      year={2023},
      eprint={2312.06838},
      archivePrefix={arXiv},
      primaryClass={hep-ex},
      url={https://arxiv.org/abs/2312.06838}, 
}

@article{Suarez_2025,
   title={Energy efficiency trends in HPC: what high-energy and astrophysicists need to know},
   volume={13},
   ISSN={2296-424X},
   url={http://dx.doi.org/10.3389/fphy.2025.1542474},
   DOI={10.3389/fphy.2025.1542474},
   journal={Frontiers in Physics},
   publisher={Frontiers Media SA},
   author={Suarez, Estela and others},
   year={2025},
   month=apr }

@misc{wegmeth2025greenrecommendersystemsunderstanding,
      title={Green Recommender Systems: Understanding and Minimizing the Carbon Footprint of AI-Powered Personalization}, 
      author={Lukas Wegmeth and Tobias Vente and Alan Said and Joeran Beel},
      year={2025},
      eprint={2509.13001},
      archivePrefix={arXiv},
      primaryClass={cs.IR},
      url={https://arxiv.org/abs/2509.13001}, 
}

@inproceedings{Pathania_2025,
  author    = {Pathania, Priyavanshi and others},
  title     = {Calculating Software’s Energy Use and Carbon Emissions: A Survey of the State of Art, Challenges, and the Way Ahead},
  booktitle = {Proceedings of the 2025 IEEE/ACM 9th International Workshop on Green and Sustainable Software (GREENS)},
  publisher = {IEEE},
  address   = {New York, NY, USA},
  year      = {2025},
  pages     = {92--99},
  doi       = {10.1109/greens66463.2025.00018}
}

@article{Caron_2026,
   title={Strategic white paper on AI infrastructure for particle, nuclear, and astroparticle physics: insights from JENA and EuCAIF},
   volume={7},
   ISSN={2632-2153},
   url={http://dx.doi.org/10.1088/2632-2153/ae35cd},
   DOI={10.1088/2632-2153/ae35cd},
   number={1},
   journal={Machine Learning: Science and Technology},
   publisher={IOP Publishing},
   author={Caron, Sascha and others},
   year={2026},
   month=jan, pages={013002} }

@article{Elvira2024SnowmassComputing,
  author  = {Elvira, V. Daniel},
  title   = {Computing within the 2021 Snowmass process},
  journal = {EPJ Web of Conferences},
  volume  = {295},
  pages   = {13003},
  year    = {2024},
  doi     = {10.1051/epjconf/202429513003}
}

@article{MUHOZA2023100930,
title = {Power consumption reduction for IoT devices thanks to Edge-AI: Application to human activity recognition},
journal = {Internet of Things},
volume = {24},
pages = {100930},
year = {2023},
issn = {2542-6605},
doi = {https://doi.org/10.1016/j.iot.2023.100930},
url = {https://www.sciencedirect.com/science/article/pii/S2542660523002536},
author = {Aimé Cedric Muhoza and Emmanuel Bergeret and Corinne Brdys and Francis Gary}
}

@misc{tu2023deepen2023energydatasetsedge,
      title={DeepEn2023: Energy Datasets for Edge Artificial Intelligence}, 
      author={Xiaolong Tu and Anik Mallik and Haoxin Wang and Jiang Xie},
      year={2023},
      eprint={2312.00103},
      archivePrefix={arXiv},
      primaryClass={cs.LG},
      url={https://arxiv.org/abs/2312.00103}, 
}

@article{Shi_2025,
   title={Satellite edge artificial intelligence with large models: architectures and technologies},
   volume={68},
   ISSN={1869-1919},
   url={http://dx.doi.org/10.1007/s11432-024-4425-y},
   DOI={10.1007/s11432-024-4425-y},
   number={7},
   journal={Science China Information Sciences},
   publisher={Springer Science and Business Media LLC},
   author={Shi, Yuanming and Zhu, Jingyang and Jiang, Chunxiao and Kuang, Linling and Letaief, Khaled Ben},
   year={2025},
   month=jun }

@misc{girgin2025edgeaidroneautonomousconstruction,
      title={EdgeAI Drone for Autonomous Construction Site Demonstrator}, 
      author={Emre Girgin and Arda Taha Candan and Coşkun Anıl Zaman},
      year={2025},
      eprint={2505.09837},
      archivePrefix={arXiv},
      primaryClass={cs.RO},
      url={https://arxiv.org/abs/2505.09837}, 
}

@article{Prabha2026EdgeAIHealthcare,
  author  = {Prabha, M. and Nandhini, S. and Dayanidhy, M. and others},
  title   = {Edge-AI integrated secure wireless IoT architecture for real time healthcare monitoring and federated anomaly detection},
  journal = {Scientific Reports},
  volume  = {16},
  pages   = {574},
  year    = {2026},
  doi     = {10.1038/s41598-025-30150-x},
  url     = {https://doi.org/10.1038/s41598-025-30150-x}
}

@misc{gonski2026machinelearningheterogeneousedge,
      title={Machine Learning on Heterogeneous, Edge, and Quantum Hardware for Particle Physics (ML-HEQUPP)}, 
      author={Julia Gonski and others},
      year={2026},
      eprint={2602.22248},
      archivePrefix={arXiv},
      primaryClass={physics.ins-det},
      url={https://arxiv.org/abs/2602.22248}, 
}

@article{UBOLDI2022166371,
title = {Extracting low energy signals from raw LArTPC waveforms using deep learning techniques — A proof of concept},
journal = {Nuclear Instruments and Methods in Physics Research Section A: Accelerators, Spectrometers, Detectors and Associated Equipment},
volume = {1028},
pages = {166371},
year = {2022},
issn = {0168-9002},
doi = {https://doi.org/10.1016/j.nima.2022.166371},
url = {https://www.sciencedirect.com/science/article/pii/S016890022200047X},
author = {Lorenzo Uboldi and others}
}

@misc{petroff2024accessiblecolorsequencesdata,
      title={Accessible Color Sequences for Data Visualization}, 
      author={Matthew A. Petroff},
      year={2024},
      eprint={2107.02270},
      archivePrefix={arXiv},
      primaryClass={cs.GR},
      url={https://arxiv.org/abs/2107.02270}, 
}

@article{Scholberg2018SupernovaMassOrdering,
  author  = {Scholberg, Kate},
  title   = {Supernova signatures of neutrino mass ordering},
  journal = {Journal of Physics G: Nuclear and Particle Physics},
  volume  = {45},
  number  = {1},
  pages   = {014002},
  year    = {2018},
  doi     = {10.1088/1361-6471/aa97be}
}

@article{DUNE2025SupernovaPointing,
  author  = {Abed Abud, A. and others},
  collaboration = {DUNE Collaboration},
  title   = {Supernova pointing capabilities of DUNE},
  journal = {Physical Review D},
  volume  = {111},
  pages   = {092006},
  year    = {2025},
  doi     = {10.1103/PhysRevD.111.092006}
}

@misc{cuesta2024supernovasolarneutrinosearches,
      title={Supernova and solar neutrino searches at DUNE}, 
      author={C. Cuesta},
      year={2024},
      eprint={2311.06134},
      archivePrefix={arXiv},
      primaryClass={hep-ex},
      url={https://arxiv.org/abs/2311.06134}, 
}

@misc{domingo2024novelprotondecaysignature,
      title={A Novel Proton Decay Signature at DUNE, JUNO, and Hyper-K}, 
      author={Florian Domingo and Herbi K. Dreiner and Dominik Köhler and Saurabh Nangia and Apoorva Shah},
      year={2024},
      eprint={2403.18502},
      archivePrefix={arXiv},
      primaryClass={hep-ph},
      url={https://arxiv.org/abs/2403.18502}, 
}

@article{Govorkova_2026,
doi = {10.1088/2632-2153/ae8218},
url = {https://doi.org/10.1088/2632-2153/ae8218},
year = {2026},
month = {jul},
publisher = {IOP Publishing},
volume = {7},
number = {4},
pages = {045024},
author = {Govorkova, Katya and García Pardiñas, Julián and Lončar, Vladimir and Schmitt, Sebastian and Nguyen, Victoria and Pizzichemi, Marco and Martinazzoli, Loris and Anne Smith, Eluned},
title = {Enabling low-latency machine learning on radiation-hard FPGAs with hls4ml},
journal = {Machine Learning: Science and Technology}
}

@ARTICLE{10735136,
  author={Sipos, Roland},
  journal={IEEE Transactions on Nuclear Science}, 
  title={The Ethernet Readout of the DUNE DAQ System}, 
  year={2025},
  volume={72},
  number={3},
  pages={317-324},
  doi={10.1109/TNS.2024.3486059}}

@misc{IntelI7,
  author       = {{Intel Corporation}},
  title        = {Intel Core i7-13650HX Processor (24M Cache, up to 4.90 GHz): Product Specifications},
  howpublished = {\url{https://www.intel.com/content/www/us/en/products/sku/232101/intel-core-i713650hx-processor-24m-cache-up-to-4-90-ghz/specifications.html}},
  note         = {Accessed 13 October 2026},
  year         = {2023}
}

@misc{NVIDIARTX4060,
  author       = {{NVIDIA Corporation}},
  title        = {GeForce RTX 4060 Family: Specifications},
  howpublished = {\url{https://www.nvidia.com/en-us/geforce/graphics-cards/40-series/rtx-4060-4060ti/}},
  note         = {Accessed 13 October 2026},
  year         = {2023}
}

@misc{NVIDIAA100,
  author       = {{NVIDIA Corporation}},
  title        = {NVIDIA A100 Tensor Core GPU Datasheet},
  howpublished = {\url{https://www.nvidia.com/content/dam/en-zz/Solutions/Data-Center/a100/pdf/nvidia-a100-datasheet-us-nvidia-1758950-r4-web.pdf}},
  note         = {Accessed 13 October 2026},
  year         = {2021}
}

@misc{CoralDatasheet,
  author       = {{Google LLC}},
  title        = {Coral USB Accelerator Datasheet},
  howpublished = {\url{https://coral.ai/docs/accelerator/datasheet/}},
  note         = {Accessed 13 October 2026},
  year         = {2019}
}

@article{Nagel2021WhitePaper,
  author  = {Nagel, Markus and Fournarakis, Marios and Amjad, Rana Ali and Bondarenko, Yelysei and van Baalen, Mart and Blankevoort, Tijmen},
  title   = {A White Paper on Neural Network Quantization},
  journal = {arXiv preprint arXiv:2106.08295},
  year    = {2021}
}

@misc{EfficientNetLite,
  author       = {Liu, Renjie},
  title        = {Higher accuracy on vision models with {EfficientNet-Lite}},
  howpublished = {TensorFlow Blog, \url{https://blog.tensorflow.org/2020/03/higher-accuracy-on-vision-models-with-efficientnet-lite.html}},
  year         = {2020}
}

@misc{CoralModelReqs,
  author       = {{Google LLC}},
  title        = {TensorFlow models on the Edge TPU},
  howpublished = {\url{https://coral.ai/docs/edgetpu/models-intro/}},
  note         = {Accessed 13 October 2026}
}

\end{document}